\documentclass[epj]{svjour}
\usepackage{graphics}
\begin{document}
\title{A critical analysis on deeply bound kaonic states in nuclei and 
the KEK experiment}
\author{E. Oset\inst{1} 
\and H. Toki\inst{2}
}                     
%
%
\institute{Departamento de F\'{\i}sica Te\'orica and IFIC,
Centro Mixto Universidad de Valencia-CSIC, 
Institutos de
Investigaci\'on de Paterna, Aptd. 22085, 46071 Valencia, Spain \and 
Research Center for Nuclear Physics, Osaka University,
Ibaraki, Osaka 567-0047, Japan}
\date{Received: date / Revised version: date}
%
\abstract{We critically discuss the theoretical developments that led to
predictions of very deeply bound kaon states and then revise the data
of the KEK experiment from where claims for evidence of deeply bound kaon 
atoms in nuclei were made. We conclude that the peaks seen in the experiment
correspond to the absorption of kaons by a pair of nucleons leading to 
$\Sigma p$ and $\Lambda p$ and leaving the daughter  nucleus as a spectator.
These conclusions have been reconfirmed by a recent experiment at FINUDA. 
}
%
\maketitle
\section{Introduction}
\label{intro}
  The success of the theoretical predictions and posterior finding of deeply 
bound pionic atoms lead to a search for other possible deeply bound states in 
nuclei.  The case of deeply bound kaons was a good candidate since from data on
kaon atoms it was known that the strong interaction of negative kaons with
nuclei was attractive. This was in spite of the fact that the low density
theorem  $\Pi = t \rho$ gives repulsion since the isoscalar $K^- N$ amplitude
at threshold is repulsive.  The consideration of Pauli blocking in intermediate
states was shown to be responsible for this change of sign in the interaction
\cite{kochpau} although further considerations of selconsistency in the
calculations led to a much weaker attraction than that produced by Pauli 
blocking alone \cite{lutz,angelsself,schaffner}. These latter works were
performed within the context of chiral unitary theory and all them led to
moderate potentials of the order of 50 MeV attraction at normal nuclear matter
density. By means of these, deeply bound $K^-$ states could be accommodated in
medium and heavy nuclei with a binding of about 30 MeV, but unfortunately a very
large width of the order of 100 MeV, which would make the spectroscopy of such
states a hopeless task. It was then a surprise that a theoretical paper 
appeared predicting a much larger kaon attraction in nuclei with three and 
four nucleons with 108 MeV binding in $^3 He$ with $I=0$  \cite{akaishi}.
Inspired by this prediction an experiment was done at KEK with $K^-$ absorption
at rest leading to proton emission. A peak in the proton momentum spectrum
was found, but if it were interpreted as corresponding to a  bound kaon atom the
binding energy would have been 195 MeV and the isospin would be $I=1$. In view
of the clear discrepancy from the predictions a paper was published
claiming the discovery of a strange tribaryon \cite{suzuki}. This experimental
finding stimulated further work and the authors of \cite{akaishi} made some
corrections to their earlier work from where they found a stronger potential 
of 618 MeV attraction in the center of the nucleus due among other rough
approximations to allowing the nucleus get compressed to {\bf ten times} the 
nuclear matter density \cite{akainew}. From that moment on the KEK work 
was presented in Conferences as an evidence for deeply bound kaons in nuclei.
  In a posterior paper \cite{prc}, the authors of the present work made a 
  critical review of
the theoretical approach of \cite{akaishi} and found a natural intepretation
for the peaks seen in \cite{suzuki}, the essence of which we present here,
together with further developments spurred by that publication.

\section{Criticism of the theoretical potential leading to deeply bound kaon
atoms}
  In \cite{prc} we make a thorough discussion of the calculations involved in
the chiral models \cite{lutz,angelsself,schaffner} which serves as a basis of
discussion for the differences and defficiencies of the work of \cite{akaishi}.
Limitation of space only allows here a brief description of these defficiencies:

1) In the chiral models a coupled channel approach is made which contains the
channels $\bar{K}N, \pi \Sigma, \pi \Lambda, \eta \Sigma, \eta \Lambda, K \Xi$.
In the work of \cite{akaishi} only the $\bar{K}N,\pi \Lambda, \pi \Sigma$ channels
 are considered but the couplings of $\pi \Lambda$ and $\pi \Sigma$ to themselves
 is neglected in spite of their large strength obtained from chiral Lagrangians.
 This restriction prevents the authors to get two $\Lambda(1405)$ states which
 are strongly supported experimentally now \cite{twolamb}. 
 
2) The authors of \cite{akaishi} assume the $\Lambda(1405)$  to be a bound state
of $\bar{K} N$, while in chiral theories the two poles are a complicated mixture
of coupled channles. These assumptions lead to a $\bar{K} N$ amplitude below
threshold that goes from a factor two  about 50 \% larger than chiral models
depending on the model.

3) One of the most serious problem in the approach of \cite{akaishi} is the lack of
selfconsistency. In the chiral calculations the $\bar{K}$ potential is obtained
and inserted in the $\bar{K}$  propagators of the loops and the procedure is
itererated till convergence is found. This is a must in the calculations because
of the presence of the $\Lambda(1405)$ resonance close to threshold. Indeed
changes in the masses of the particles move the resonance up and down and 
one goes easily from attraction to repulsion of the $\bar{K}$. Needless to say
 that with a potential of 618 MeV attraction as in \cite{akainew} the
selfconsistent consideration its absolutely necessary, but it is not done in 
\cite{akainew}.  This lack is sufficient too make the results unreliable.

4) Since not enough attraction is obtained to get deep and narrow states the
nuclear matter is compressed to {\bf ten times} nuclear matter density in 
\cite{akaishi}. We consider this utterly exagerated. 

\label{sec:1}

\section{Discussion on the KEK experiment}
\label{sec:1}
On the basis of the discussion in the former section it is quite clear that 
the KEK experiment \cite{suzuki} could not be
interpreted in terms of the creation of deeply bound kaonic states. From
this perspective we would like to interpret the meaning of the peak seen there.
The experiment is 
\begin{equation}
stopped~ K^- +~^4He \to S +~p
\label{triprod}
\end{equation}
where S has a mass of about 3115 MeV, and has the quantum numbers
of $YNN$ with zero charge where $Y$ is a S=-1 hyperon. The state S
would have $I_3=-1$ and hence cannot be $I=0$ as predicted
originally in Ref. \cite{akaishi}, which was already noted in
Ref. \cite{suzuki}. The peak is also quite narrow, around or smaller
than 20 MeV. In \cite{prc} a discussion is made of possible mechanisms producing
the peak and all them are eliminated for inconsistencies with other data. Only
one possible reaction passed all the experimental tests and we describe it
below.

In view of the previous unsuccessful trials it was concluded in 
\cite{prc} that the
reaction which is most likely to happen is  $K^-$
absorption by two nucleons in $^4$He leaving the other two
nucleons as spectators. This kaon absorption process should happen
from some $K^-$ atomic orbits which overlap with
the tail of the nuclear density and hence the Fermi motion of the
nucleons is small.  Then we would have $K^- NN \to \Lambda N$,
and $\Sigma N$, and the two baryons are emitted back to back with
the momentum for the proton of 562, and 488 MeV/c respectively.
These results are very interesting: the peak of the proton
momentum in Ref. \cite{suzuki}, before  proton energy loss corrections,
appears at 475 MeV/c (see Fig. 5 of this reference). This matches
well with the 488 MeV/c proton momentum from a  $K^- NN \to \Sigma
N$ event, and the proton would  lose about 13 MeV/c when crossing
the thick target. This energy loss is compatible with the
estimate of about 30 MeV/c in Ref. \cite{suzuki}, particularly taking
the width of the peak also into account.

This suggestion sounds good, but then one could ask oneself:  what
about $K^- NN \to \Lambda N$? Should not there be another peak
around 550 MeV/c, counting also the energy loss? The logical
answer is yes, and curiously one sees a second peak around 545
MeV/c in the experiment. The peak is clearly visible although less
pronounced than the one at 475 MeV/c and it appears in the region
of fast decline of the cross section.

There are other arguments supporting our suggestion. Indeed, as
mentioned above, the pion momenta from $\Lambda$ decay are smaller 
than those from $\Sigma$ decay. As a consequence of this
we should expect the peak associated with $p \Lambda$ emission to
appear in the low momentum side of the pion (the range of pion momenta is from
61 MeV/c to 146 MeV/c from phase space considerations). Actually, this is the
case in the experiment of Ref. \cite{suzuki} as one can see in Fig. 5d
of this reference, corresponding to the spectrum when the low pion cut is
applied, where the peak of higher momentum stands out more clearly.
 On the other hand, by working out the phase space for
$\Sigma$ decay, the pion momenta range from 162 Mev/c to 217 MeV/c and the pion 
could be
seen in the two regions of pion momenta  of Ref. \cite{suzuki}, as it is indeed 
the case (see figs. 5c, 5d of Ref. \cite{suzuki}). 

One can even argue about the size of the peaks and their relative
strength. For this the information of
Ref. \cite{katz} is very useful. There we find the following results
for events per stopped $K^-$:

\begin{eqnarray}
\Sigma^- p~d ~~~~~1.6 ~\%
\label{sigma1}\\
\Sigma^- ppn  ~~~~2.0 ~\%
\label{sigma2}\\
\Lambda (\Sigma^0) pnn ~~~~11.7 ~\%
\label{sigma3}
\end{eqnarray}
with errors of 30-40 \%.

From these data and further anaylsis is was concluded in \cite{prc} that one
could understand the larger strength of the peak coming from $\Sigma p$ versus
that from $\Lambda p$.  It was also suggested that because the $\Sigma^- p~d$
reaction leaves indeed the daugher nucleus ($d$ ) untouched, this channel is the
best candidate to explain the peak caused by the $\Sigma p$ emission, leading
to a momentum of the proton of 482 MeV/c.

The hypothesis advanced should have other consequences.  Indeed,
this peak should not be exclusive of the small nuclei. This should
happen for other nuclei. Actually in other nuclei, let us say $K^-
~ ^7$Li, the signals that we are searching for should appear when
a proton  as well as a $\Sigma$ or a $\Lambda$ would be emitted
back to back and a residual nuclear system remains as a spectator and
stays nearly in its ground state.
We would thus expect
two new features: first, the two peaks should be there. However,
since now the spectator nuclear systems remain nearly in their
ground states, only about the
binding energy of the two participant nucleons will have to be
taken from the kaon mass, instead of the 28 MeV in $^4$He for a
full break up, as a consequence of which the proton momenta should
be a little bigger. We make easy estimates of 502 MeV/c for the
proton momentum in the case of $p \Sigma$ emission and 574 MeV/c
for the case of $p \Lambda$ emission.  Curiously the FINUDA data
\cite{piano} exhibits two peaks in $^7 Li$ around 505 MeV/c and
570 MeV/c (see comments at the end about a recent publication).

There is one more prediction we can make. The process discussed
has to leave the remnant nucleus in nearly its ground state.
This means
that one has to ensure that the nucleus is not broken, or excited
largely, when the energetic protons go out of the nuclear system.
Theoretically one devises this in
terms of a distortion factor that removes events when some
collision  of the particles with the nucleons takes place.
Obviously this distortion factor would reduce the cross sections
more for heavier nuclei and, hence, we should expect the signals
to fade away gradually as the nuclear mass number increases. This
is indeed a feature of the FINUDA data \cite{piano}.

 Similarly, we can also argue that the spectator nucleus, with a momentum equal 
 to that of the combined pair on which $K^-$ absorption occurs, will have smaller
 energies for heavier nuclei since their mass is larger. Hence, the spreading of
 the energy of the emitted proton should become narrower for heavier nuclei. 
 This is indeed a feature of the FINUDA data \cite{piano}.

This sequence of predictions of our  hypothesis,
confirmed by the data of \cite{suzuki} and \cite{piano}, provides
a strong support for the mechanism suggested of $K^-$ absorption
by a pair of nucleons leaving the rest of the nucleons as
spectators. Certainly, further tests to  support this idea,
or eventually refute it, should be  welcome. An
obvious test is to search for $\Sigma$ or $\Lambda$ in
coincidence and correlated back to back with the protons of the peak.

\section{Conclusions}
In this paper we have made a thorough review of the theoretical
developments that led to predictions of deeply bound kaonic atoms
in light nuclei.  We could show that there were many
approximations done, which produced unreliably deep potentials.  Three
 main reasons made the
approximations  fail dramatically:  the problem of the coupled
channels to produce the two $\Lambda(1405)$ states was reduced to
only one channel, the $\bar{K}N$, and only one $\Lambda(1405)$,
 which was assumed to be
a bound state of the $\bar{K}N$ potential was considered.   
The second serious problem was the lack of
selfconsistency in the intermediate states, which makes the
results absolutely unreliable when one is in the vicinity of a
resonance, as is the case here. The last serious problem is allowing 
the matter to be compressed to {\bf ten time } nuclear matter density.
Altogether one obtains a
potential as large as ten times what one gets without making these
approximations.  We also state  that the width becomes zero in the
I=0 channel due to the binding energy being lower than the
pion-$\Sigma$ threshold.  However, the selfconsistency
consideration automatically produces the two body kaon absorption
processes, which provides large widths. 

With the weakness of the theoretical basis exposed and the realization that
binding energies of 200 MeV for a kaon in a system of three
particles are out of scale, we looked for a plausible
explanation of present experiments which could be interpreted
differently than creating these deeply bound $K^-$ states.
After using information and the analysis of Ref. \cite{suzuki} discarding
potential alternatives with chain reactions, we were led by
elimination to a  source of explanation deceivingly simple, and
which, however,  passes the present experimental tests: the
association of the observed proton peaks to $K^-$ absorption by a
nucleon pair leaving the rest of the nucleons as spectators. From
the emission of $p \Sigma$ we explained the peak found in
Ref. \cite{suzuki} and from the emission of $p \Lambda$ we predicted
another peak which is indeed present in the experiment of
Ref. \cite{suzuki}.  The hypothesis made led us to conclude that these
peaks should also be visible in other nuclei at slightly larger
proton momenta, should be narrower and their strength should
diminish with increasing mass number of the nucleus. Fortunately,
all these predictions could be tested with present data from
FINUDA \cite{piano} and all the predictions were confirmed by
these  data.

The FINUDA data quoted in this work as \cite{piano} have been 
recently published
\cite{lastfinuda} and there they reconfirm our claims for the interpretation of
these peaks, that are seen in other light nuclei and tend to disappear in 
heavy ones, as we predicted. 

\section{Acknowledgments}
 This work is 
partly supported by DGICYT contract number BFM2003-00856, the Generalitat
Valenciana, the CSIC-JSPS collaboration agreement
and the E.U. EURIDICE network contract no. HPRN-CT-2002-00311.
This research is part of the EU Integrated Infrastructure Initiative
Hadron Physics Project under contract number RII3-CT-2004-506078.

\end{document}